# Multiple imputation for longitudinal data: A tutorial

**Authors:** Rushani Wijesuriya [1,2], Margarita Moreno-Betancur[1,2,] John B Carlin [1,2], Ian R White[3], Matteo Quartagno[3] and Katherine J Lee [1,2]
1 Clinical Epidemiology & Biostatistics (CEBU), Murdoch Children's Research Institute
2 Department of Paediatrics, University of Melbourne
3 MRC Clinical Trials Unit, Institute for Clinical Trials and Methodology, University College London, London, United Kingdom

## Abstract

Longitudinal studies are frequently used in medical research and involve collecting repeated measures on individuals over time. Observations from the same individual are invariably correlated and thus an analytic approach that accounts for this clustering by individual is required. While almost all research suffers from missing data, this can be particularly problematic in longitudinal studies as participation often becomes harder to maintain over time. Multiple imputation (MI) is widely used to handle missing data in such studies. When using MI, it is important that the imputation model is compatible with the proposed analysis model. In a longitudinal analysis, this implies that the clustering considered in the analysis model should be reflected in the imputation process. Several MI approaches have been proposed to impute incomplete longitudinal data, such as treating repeated measurements of the same variable as distinct variables or using generalized linear mixed imputation models. However, the uptake of these methods has been limited, as they require additional data manipulation and use of advanced imputation procedures. In this tutorial, we review the available MI approaches that can be used for handling incomplete longitudinal data, including where individuals are clustered within higher-level clusters. We illustrate implementation with replicable R and Stata code using a case study from the Childhood to Adolescence Transition Study.

## 1. Introduction

In longitudinal studies, data are collected from the same participants over time, often at semi-regular time intervals referred to as waves. In such studies, repeated observations from the same individual will be correlated with each other, resulting in non-independence of observations.[1] These correlations should be appropriately accounted for in analysis, as failure to do so can result in incorrect variance estimates.[1]

While almost all research studies suffer from missing data, this issue can be particularly prominent in longitudinal studies as there are multiple opportunities for missing data. Missing data can lead to biased inferences and reduced precision if not handled appropriately.[2, 3] Multiple imputation (MI) has become increasingly popular for handling missing data, including in longitudinal studies.[4] Originally proposed by Rubin, MI involves two key steps.[5]  First, imputations for the missing values are drawn from their predictive distribution given the observed data using a model fitted to the observed data. This process is carried out multiple (*m*) times to produce multiple (*m*) completed datasets. Next, each of the completed datasets is analysed using the substantive model of interest and the resulting estimates and their standard errors combined using Rubin's rules to obtain the final MI estimate of the parameter(s) of interest.[6]

Joint modelling (JM) and fully conditional specification (FCS) are the two most common frameworks for conducting MI when there is missingness in multiple variables.[7, 8] The JM approach imputes all variables with missing values simultaneously by assuming a joint multivariate model for these variables, while the FCS approach imputes incomplete variables sequentially by using a series of univariate imputation models for each incomplete variable in turn.[7] Standard implementations of JM and FCS , if appropriately applied, are warranted to provide unbiased estimation of the parameters of interest if the propensity for missing data is independent of the missing data conditionally on the observed data, referred to as the missing at random (MAR) assumption, but can also provide valid inferences in other contexts, as detailed elsewhere.[9]

The validity of results from MI is contingent on having an appropriately specified imputation model that is compatible with the substantive analysis model.[10] In principle, compatibility means that there exists a joint model whose conditional distributions include both the imputation model and the substantive analysis model. In practice, compatibility is achieved by tailoring the imputation model to include features of the substantive analysis i.e., all variables in the same form as in the analysis model, including any interactions and non-linear terms.[11] In the context of a longitudinal analysis that models clustering by individual, it is important that the imputation model also accounts for this clustering.[3, 12-14] Importantly, an imputation model can be more general than the substantive model and include additional variables that are not in the analysis model, referred to as auxiliary variables, which can be used to reduce bias or improve precision.[15, 16] In the context of longitudinal data, an imputation model can exploit auxiliary information from other waves to impute values at a given wave.

A number of MI approaches have been proposed for accounting for correlated data in the imputation model, many of which can be implemented in standard software such as R and Stata,[17-19] but their application requires additional data manipulation and/or use of specialized procedures.[20] While several studies have reviewed and evaluated these approaches by way of simulation studies,[14, 21-23] the available guidance on their implementation is limited. In this tutorial, we provide an overview of the available MI approaches that can be used for imputing incomplete longitudinal data, including where longitudinal data are further clustered within higher-level clusters, for example schools or geographical areas. We demonstrate the implementation of these approaches with replicable code in R using an empirical example based on Childhood to Adolescence Transition Study (CATS). Comparable code for Stata is provided in a GitHub repository where available.[24]

The structure of the paper is as follows. In section 2, we describe CATS case study, which uses a random intercept linear mixed model (LMM) as the substantive analysis model. In section 3 we review and illustrate a range of MI approaches for handling incomplete longitudinal data. In section 4, we review extensions of the MI approaches to situations where there is clustering of the participants within higher-level clusters. Section 5 briefly demonstrates, using CATS, considerations for choosing a method in practice amongst those reviewed. We then end with a brief discussion in section 6.

## 2. Case study: Childhood to Adolescence Transition Study (CATS)

CATS is a longitudinal population-based cohort study based in Melbourne, Australia, with the broad aim of understanding the factors that influence young peoples' health and emotional wellbeing as they transition through their teenage years.[25] The study recruited 1239 grade 3 students (8-9 years) from a stratified (government, catholic and independent schools) random sample of 43 schools in 2012.[25] The students have been followed up annually using parent, teacher, and student self-report questionnaires. The full details of the study can be found elsewhere.[25] In this tutorial, we focus on an analysis that assesses the effect of depressive symptoms (at waves 2, 4 and 6) on subsequent academic performance (at waves 3, 5 and 7) measured using numeracy scores obtained for the National Assessment Program – Literacy and Numeracy (NAPLAN), which is a test administered to all students in schools across Australia. In estimating this effect, the analysis adjusts for potential confounders at wave 1: NAPLAN numeracy score, sex, socio-economic status (SES), and age. To account for the correlations among repeated measures, our substantive model is a random intercept LMM as shown in (1), where we include a random effect (intercept) to allow for the correlation among repeated measures within an individual:

$$\text{numeracy.score}_{jk} = \beta_0 + \beta_1 * \text{prev.dep}_{jk} + \beta_2 * k + \beta_3 * \text{numeracy.score}_{j1} \quad (1)$$
$$+\beta_4 * \text{sex}_j + \sum_{a=1}^{4} \beta_{5,a} * I\left[\text{SES}_{j1} = a\right] + \beta_6 * \text{age}_{j1} + b_{0j} + \varepsilon_{jk}$$

where $j$ denotes the $j^{th}$ individual, $k$ denotes the $k^{th}$ wave ($k = 3,5,7$), I[.] denotes a binary indicator function, $\varepsilon_{jk}$ denotes independent random measurement errors distributed as $N(0, \sigma_e^2)$ and $b_{0j}$ denotes individual-level random effects distributed as $N(0, \sigma_b^2)$. The rest of the notation is described in Table 1, where the proportions with missing values for each variable are also shown.

In (1), $\beta_1$ represents the estimand of interest, the causal effect of early depression on academic performance, which for simplicity is assumed to be constant over time (as well as constant across strata of the confounders, as usual in outcome regression estimation), although in practice this can be relaxed. CATS data are not publicly available due to ethical requirements. Instead, throughout the tutorial we use a simulated dataset that closely mimics the actual data. The data simulation process, including the directed acyclic graph (DAG) encapsulating the assumed causal dependencies among the variables, is explained in detail in the supplementary files (sections S2 and S3). The code for simulating the data can be found at https://github.com/rushwije/Longitudinal_multiple_imputation_tutorial.

Most statistical methods for analysing longitudinal data require the data to be in long format, where there is one row per individual per time point (Figure 1a). With the data in long format, one could fit the LMM in (1) using an available case analysis (ACA), which ignores records with missing values in any of the variables in the model. In CATS example, the ACA includes only 54% (1943/3600) of the records in the long-format dataset. The estimate for the causal effect of interest, $\beta_1$, from ACA in CATS, along with its standard error and the variance component estimates, are given in the supplementary files: section S1.

Longitudinal data can also be arranged into wide format if the repeated measures occur at aligned time-points (e.g. waves) across individuals (i.e., data are balanced), as shown in Figure 1b. In practice, data are sometimes analysed in wide format for convenience in calculating summary measures across individuals.

Transforming data from long to wide and wide to long formats can be done easily in R using the reshape() function:

```r
#Reshaping CATS_long (in long format) to wide format
CATS_wide <- reshape(CATS_long, v.names = c("numeracy_score", "prev_dep", "prev_sdq"),
    timevar = "time", idvar = "id", direction = "wide")

#Reshaping CATS_wide (in wide format) to long format
CATS_long <- reshape(CATS_wide, varying = list(c("prev_dep.3", "prev_dep.5", "prev_dep.7"),
    c("numeracy_score.3", "numeracy_score.5", "numeracy_score.7"), c("prev_sdq.3",
      "prev_sdq.5", "prev_sdq.7")), idvar = "id", v.names = c("prev_dep", "numeracy_score",
    "prev_sdq"), times = c(3, 5, 7), direction = "long")
```

```
a) head(round(CATS_long),10)
   school id age sex ses numeracy_scoreW1 time prev_dep numeracy_score prev_sdq
        1  1   8   1   5                2    3        1              2       22
        1  1   8   1   5                2    5        1              2       20
        1  1   8   1   5                2    7        1              2       14
        1  2   7   0   3                0    3        1             -1       21
        1  2   7   0   3                0    5        1             -1       13
        1  2   7   0   3                0    7        1              0       16
        1  3   9   0   2               -1    3        2             -2       18
        1  3   9   0   2               -1    5        2             -2       14
        1  3   9   0   2               -1    7        2             -2       16
        1  4  10   1   1               NA    3        1             NA       18

b) head(round(CATS_wide),10)
   school id age sex ses numeracy_scoreW1 numeracy_score.3 prev_dep.3 prev_sdq.3 numeracy_score.5 prev_dep.5 prev_sdq.5
        1  1   8   1   5                2                2          1         22                2          1         20
        1  2   7   0   3                0               -1          1         21               -1          1         13
        1  3   9   0   2               -1               -2          2         18               -2          2         14
        1  4  10   1   1               NA               NA          1         18               NA          2         19
        1  5   8   0   3               -1               -1          1         14               -1          1         21
        1  6   9   1   3                0                0         NA         14                1          1         17
        1  7   9   0   5                1                1         NA         14                1          1         15
        1  8   9   0   5               -1               -2          1         15               -2          1         13
        1  9   9   0   5                0               -1          1          7                0          1         10
        1 10   8   1   5                0                1          1         15                0          1         15
```

Figure 1: CATS data in a) long format, b) wide format. Box indicates the participant with id=1.

**Insert Table 1

### 3. MI approaches for longitudinal data

There are two broad approaches to imputing longitudinal data. Firstly, longitudinal data can be imputed in wide format by imputing the repeated measures as distinct variables in the imputation model. This approach allows an unstructured (unrestricted) correlation matrix among the repeated measures. Alternatively, longitudinal data can be imputed using LMM-based MI approaches that impute repeated measures in long format and model the correlation among incomplete repeated measures via random effects, which imposes more parametric assumption on the correlation structure between repeated measures. Note that auxiliary variables, whether repeated or fixed, or complete or incomplete, are handled the same as other (analysis) variables of the same nature. We describe and illustrate both of these approaches below. We give each method an abbreviated name describing the imputation approach (JM/FCS), the number of clustering levels the MI implementation was developed to handle (1L= 1 level/2L=2 levels), and the format of the data used in the procedure (wide/long).

The current practical recommendation is that the number of imputations should be at least as large as the percentage of incomplete cases.[26] In CATS data, 46% of the records were missing in long format, while 65% of the records were missing in wide format. Therefore, we generate 66 imputations for all approaches.

### 3.1 MI approaches for imputing in wide format

### 3.1.1 Standard JM (JM-1L-wide)

JM imputation assumes that all incomplete variables jointly follow a multivariate (MV) distribution, conditional on any complete covariates,[8] most commonly a multivariate normal (MVN) distribution. Repeated measures can be imputed by applying MVN imputation to the data in wide format, which allows an unrestricted correlation structure between the variables via the unstructured covariance matrix of the assumed MVN model.

The MVN model treats all incomplete variables as continuous variables, including those that are binary or categorical, and generates imputations for each variable on a continuous scale. This is the case in many software implementations (e.g., Stata and SAS). Although it is generally recommended to use unrounded values in the target analysis where possible, if the original categories of a variable are required for the substantive analysis, the imputed values need to be categorised into its original form. A number of post-imputation rounding approaches for binary and categorical variables have been proposed[27-29], although none of them are optimal as rounding can introduce bias.[30, 31] In CATS example, the exposure (previous wave depressive symptoms) and one of the confounders (SES at age 1) are incomplete categorical variables. These variables do not require rounding for the target analysis and hence we simply use the unrounded values in the substantive analysis. However, we provide Stata syntax in GitHub for so-called adaptive rounding of binary variables, and continuous calibration rounding of ordinal categorical variables.

Another approach for imputing categorical variables (say with $K$ categories) within the MVN JM imputation model is to assume that the categorical variable arises from a set of $(K-1)$ underlying latent normal variables that are modelled as jointly normal in the MVN imputation procedure alongside the other incomplete variables.[32] The imputed values of the latent variable(s) are then transformed back to the original categorical variable. For example, for a binary variable $Y_j$, with an assumed underlying latent normal variable $Y_j^*$, $Y_j = 0$ if $Y_j^* > 0$ and $Y_j = 1$ if $Y_j^* \leq 0$. The same reasoning extends naturally to a $K$-level categorical variable $X_j$ with a set of $(K-1)$ underlying latent normal variables, say $X_j^{1*}, \ldots X_j^{(k-1)*}$, whereby $X_j = 1$ if $X_j^{1*} = \max_k X_j^{k*}$ and $X_j^{1*} > 0$; $X_j = 2$ if $X_j^{2*} = \max_k X_j^{k*}$ and $X_j^{2*} > 0$ and so on. Note that for ordinal categorical variables this approach ignores the ordered nature of the categories. This latent normal approach can be implemented in the R package **jomo,** and has been shown to perform well in simulation studies.[33] An alternative approach is to assume a single latent normal variable with multiple cut points relating to the different categories of the variable. This should be more efficient when dealing with ordinal data, as it explicitly reflects the order between categories. However currently there are no implementations of this in mainstream software. Further technical details of the two latent normal formulations can be found in elsewhere.[3, 32] Within the JM framework, the latent normal formulation for handling categorical variables is preferable over other post-imputation rounding approaches as it enables transformation between the imputed variable and the original variable within the imputation procedure.

To illustrate the JM-1L-wide approach, we use the latent normal formulation in **jomo**. The first step is that all binary and categorical incomplete variables need to be stored as type "factor"; variables stored as "numeric" will be imputed as continuous:

```
cat.vars <- c("ses", "prev_dep.3", "prev_dep.5", "prev_dep.7")
CATS_wide[cat.vars] <- lapply(CATS_wide[cat.vars], factor)
```

Next, two separate data frames need to be created: (1) incomplete variables to be imputed which will be specified as outcomes for the imputation model, and (2) complete variables to be used as predictors in the imputation model, with further inclusion a column of 1's which represents the intercept term. Complete categorical predictors if any must be included as indicator variables:

```
dataw_inc <- CATS_wide[, substr(names(CATS_wide), 1, 6) %in% c("prev_d", "numera",
    "ses")]

dataw_comp <- cbind(Intercept = rep(1, nrow(CATS_wide)), CATS_wide[,
substr(names(CATS_wide), 1, 6) %in% c("age", "sex", "prev_s")])
```

In JM, imputations for the incomplete variables are sampled using an iterative Gibbs sampling algorithm, a type of Markov Chain Monte Carlo (MCMC) algorithm.[34] The algorithm alternately samples missing values for the incomplete variables and the parameter values of the MVN model from the relevant conditional distribution until the chain converges to a stationary distribution. These initial iterations of the algorithm are referred to as the "burn-in" iterations. If the latent formulation is used, the Gibbs sampler is augmented with an additional Metropolis Hastings step to ensure identifiability of the MVN model. See [17] for a more detailed discussion of the algorithm. Once convergence is achieved, the current draw of missing values is used as the first imputed dataset. Subsequent imputed datasets are created by sampling the chain after a sufficient number of iterations to ensure that the draws are independent.[8] Therefore, before sampling the imputations, it is important to assess if the MCMC sampler has converged. This can be done using the wrapper function `jomo.MCMCchain()`, which runs the MCMC sampler for a number of iterations as specified by the `nburn` option without creating any imputations and stores estimates for all parameters of the imputation model (regression coefficients as well as the covariance matrices) at each iteration. The wrapper function automatically selects a sub-function for the sampler, depending on the model specification (e.g., not clustered or clustered) and the variable type being imputed (e.g., continuous variables only, categorical variables only or a mixture of both). In applying JM-1L-wide to the CATS example, the subfunction `jomo1mix.MCMCchain()` is automatically selected upon calling `jomo.MCMCchain()`. All available sub-functions and their details are summarised in the supplementary files.[35] To reproduce results, the random number seed is set using the `set.seed()` function:

```
set.seed(2946)
impCheck <- jomo.MCMCchain(Y = dataw_inc, X = dataw_comp, nburn = 1000)
```

The estimates for the parameters can be plotted against iteration to visually assess convergence (trace plot):

```
plot(c(1:1000), impCheck$collectbeta[1, 1, 1:1000], type = "l", ylab =
expression(beta["0,1"]), xlab = "Iteration number")
```

In Figure 2 we show the trace plot in CATS example for estimates of one of the intercepts in the imputation model (the one for the standardized numeracy score at wave 1). The trace plot should show random scatter around the mode of the distribution and not exhibit clear trends. In Figure 2, while some mild autocorrelation is observed, the estimate appears to fluctuate over a small range after approximately 900 burn-in iterations, so we use 1000 burn in iterations for the imputation model. `Jomo.MCMCchain()` should always be used before `jomo()`, but for brevity we omit this for the rest of the article.

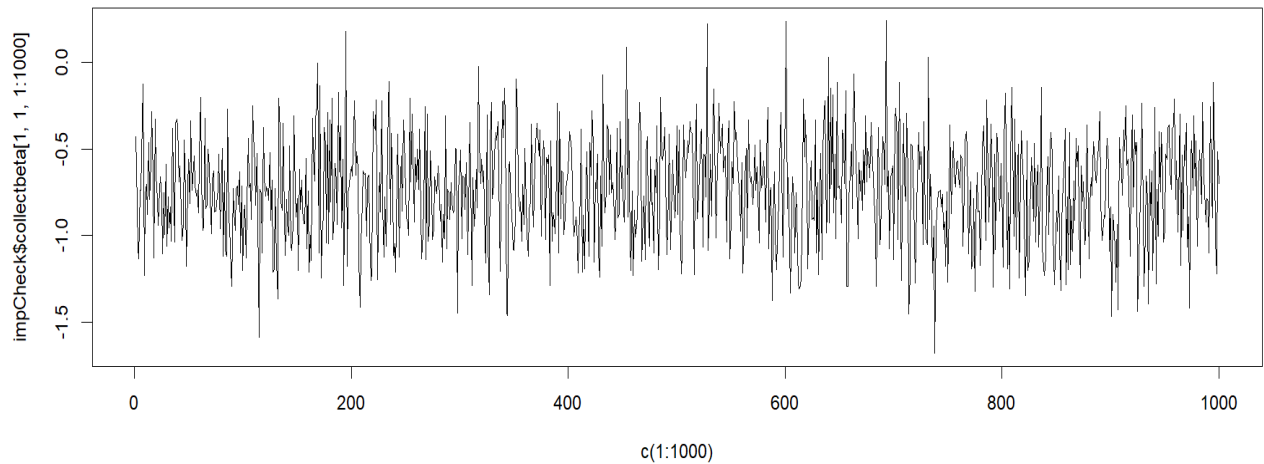

Figure 2: Trace plot for the intercept estimate corresponding to the standardized numeracy score at wave 1

Once the required burn-in iterations have been determined, the imputation model can be fitted using the `jomo()` wrapper function. The number of initial iterations in the chain is set using the `nburn` option. The between imputation iterations is set using the `nbetween` option. The default value for between imputation iterations is 1000, and we use the same. Of note, to ensure independence between imputed datasets, one could use a trace plot like that in Figure 2 to examine autocorrelation between imputations to inform this choice. Note this is not currently in jomo package but is possible in other packages (e.g. mitml, see below). The command below was used to generate the imputations using JM-1L-wide in **jomo** for CATS example:

```
set.seed(2946)
imp1 <- jomo(Y = dataw_inc, X = dataw_comp, nimp = 66, nburn = 1000, nbetween = 1000)
```

Next, the imputed datasets need to be extracted and stored in a list, converted to long format, and the substantive LMM analysis model needs to be fitted to them. This is achieved using the following code:

```
imp.list <- imputationList(split(imp1, imp1$Imputation)[-1])

imp.long <- lapply(imp.list$imputations, function(d) {
    reshape(d, varying = list(c("prev_dep.3", "prev_dep.5", "prev_dep.7"),
c("numeracy_score.3", "numeracy_score.5", "numeracy_score.7"), c("prev_sdq.3", "prev_sdq.5",
"prev_sdq.7")), idvar = "id", v.names = c("prev_dep", "numeracy_score", "prev_sdq"), times =
c(3, 5, 7), direction = "long")
})

fit.imp1 <- lapply(imp.long, function(d) {
    lmer(numeracy_score ~ prev_dep + time + age + numeracy_scoreW1 + sex + factor(ses) +
        (1 | id), data = d)
})
```

Finally, pooling of estimates from the multiple analyses is achieved using the `testEstimates()` function from the **mitml** package in R:[36]

```
testEstimates(fit.imp1, extra.pars = TRUE)
```

Of note, the **mitml** provides an alternative interface for fitting joint modelling imputation models using **jomo,** and includes additional tools for managing and visualising multiply imputed data sets, as well as examining autocorrelations between imputations.[17]

*3.1.2 Standard FCS (FCS-1L-wide)*

The FCS approach involves specifying a series of univariate conditional regression models, one for each incomplete variable conditional on all of the other variables, which can again be applied to the repeated measures in wide format. Each univariate conditional model specified can be tailored to the variable of interest. For example, linear regression can be used to impute continuous variables, logistic regression to impute binary variables, and ordinal regression to impute ordinal variables.[26, 37]

The FCS approach can be implemented via the R package **mice**, which has a range of built-in univariate imputation methods for imputing different types of variables.[7] Further details of these can be found elsewhere.[38] As with JM-1L-wide, the first step is that each variable needs to be stored as the correct type (continuous- "numeric"; categorical/binary- "factor") before specifying the imputation method (as illustrated above).

When using FCS, we must first specify the imputation model to be used for each incomplete variable. To do this, a method vector is created which specifies the different imputation methods used for imputing each incomplete variable.

```
Meth1 <- make.method(CATS_wide)
```

When creating the above vector, by default the software will fill in imputation methods for the incomplete variables (e.g., predictive mean matching (PMM)[39] which samples imputations from the observed values of the same variable for continuous variables, and logistic regression for binary variables, etc.), but this can be modified by the user as shown below. For CATS case study, we use linear regression (`norm`) for continuous, logistic regression (`logreg`) for binary, and ordinal logistic regression (also known as a proportional odds model, `polr`) for ordered categorical variables, respectively:

```
meth1[substr(names(CATS_wide), 1, 6) %in% c("prev_d")] = "logreg"
meth1[substr(names(CATS_wide), 1, 6) %in% c("numera")] = "norm"
meth1[substr(names(CATS_wide), 1, 5) %in% c("c_ses")] = "polr"
```

The predictors to be used in each univariate model are specified via a single square "predictor matrix", of dimension equal to the number of variables in the dataset, with both rows and columns representing the variables in the order in which they appear in the dataset. A cell with a "1" indicates that the column variable is included as a predictor in the imputation model for the row variable and "0" indicates that the column variable is omitted. The predictor matrix can be created using the function `make.predictorMatrix()`. By default, the created matrix sets all values, except for the diagonal values, to 1, which means that for each incomplete variable all other variables are used as predictors. However different sets of predictors may be specified by the user by updating the matrix. In the case study we create the default predictor matrix and set the cluster variables (child's ID and the school variables) to "0" in each row to omit them from the predictor sets for all variables. This can be achieved using the following code:

```
pred1 <- make.predictorMatrix(CATS_wide)
pred1[, c("id", "school")] = 0
```

Within the FCS approach, imputations are drawn for each variable in turn, cycling through each of the incomplete variables, and repeating the procedure for a specified number of iterations (t=1, …, T) until convergence, at which time the last imputed dataset is retained. This can be achieved in R using the mice function, using the `maxit` option to set the number of iterations. The whole process is repeated *m* times, with the random number seed set using `set.seed()` function:

```
set.seed(3726)
imp2 = mice(data = CATS_wide, m = 66, predictorMatrix = pred1, method = meth1, maxit = 10)
```

Monitoring convergence in **mice** is done via trace plots, using the plot() function which plots one or more statistics of interest, such as the mean and standard deviation, for imputed variables across imputations. In most practical applications of FCS, a low number of iterations (between 5-20) is usually enough to achieve convergence.[40] Figure 3 shows a trace plot of the mean and standard deviation for depressive symptoms at waves 3,5 and 7 from the case study where the streams from different imputations appear to be free of trends and intermingling well indicating convergence.

```
Plot(imp2, c("prev_dep.3", "prev_dep.5", "prev_dep.7"))
```

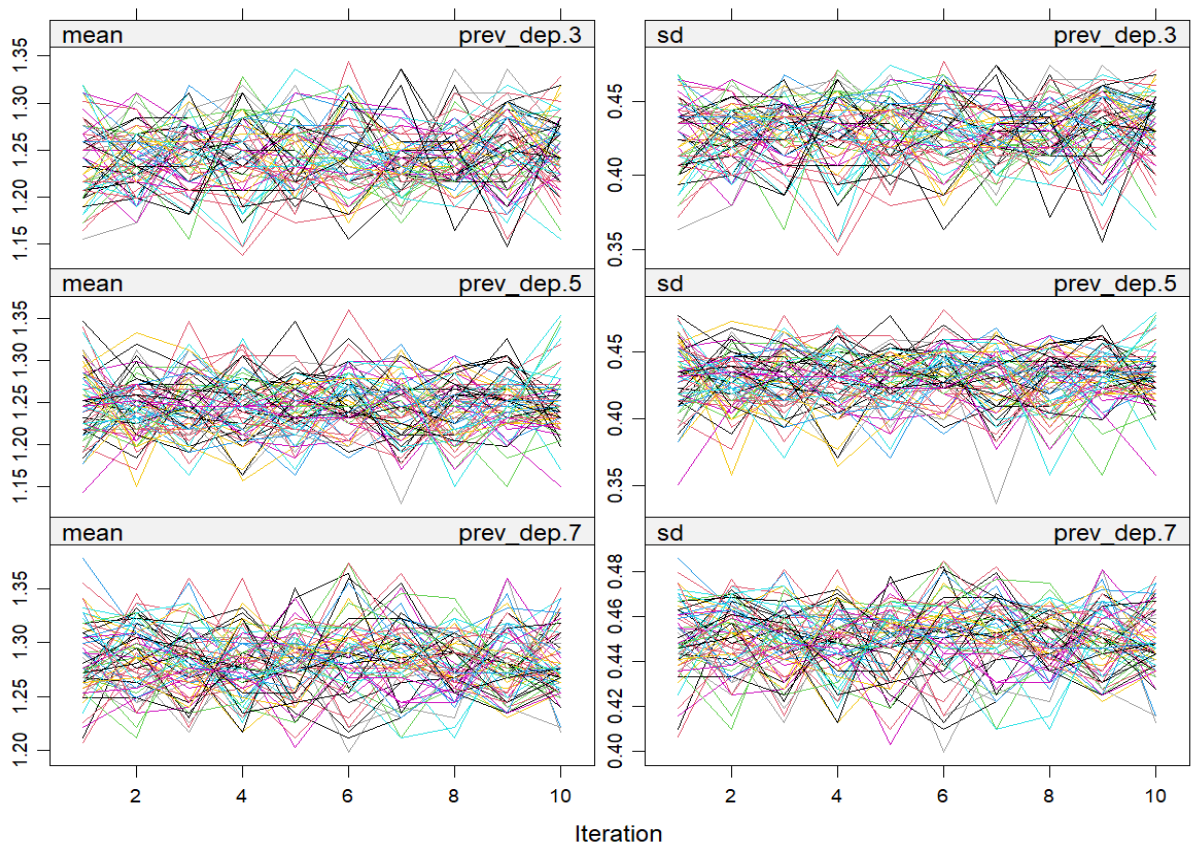

Figure 3: Trace plots for depressive symptoms at waves 3,5 and 7

As with JM-1L-wide, the imputed datasets need to be extracted and stored in a list to be reshaped to long format prior to fitting the substantive analysis model. The extraction is achieved using the code below, while the same code as used with JM-1L-wide can be used to reshape and conduct the analysis.

```
imp.list <- complete(imp2, "all")
```

### 3.1.3 Standard FCS with moving time window (FCS-1L-wide-MTW)

A variant of the FCS approach is to use a moving time window (MTW). This approach imputes each incomplete repeated measure using a subset of the time-dependent variables based on a pre-specified time window (e.g. the immediately adjacent time points). Restricting the time window in this way makes the assumption that measures being imputed are conditionally independent of the variables outside the time window given .[41] The MTW approach can be conducted by modifying the predictor matrix to omit variables outside the time window from the predictor set (i.e. by setting the relevant values to "0") in the FCS algorthm above. This is illustrated for the case study below, where we restrict the window so that only measurements at timepoints (t-1) and (t+1) are used as predictors, with the rest of the steps remaining the same as for FCS-1L-wide (see section 3.1.2).

```
pred2 <- make.predictorMatrix(CATS_wide)
pred2[, c("id", "school")] = 0
pred2["numeracy_scoreW1", c(grep("5", names(CATS_wide)), grep("7", names(CATS_wide)))] = 0
pred2[c("prev_dep.3", "prev_sdq.3", "numeracy_score.3"), grep("7", names(CATS_wide))] = 0
pred2[c("prev_dep.5", "prev_sdq.5", "numeracy_score.5"), "numeracy_scoreW1"] = 0
pred2[c("prev_dep.7", "prev_sdq.7", "numeracy_score.7"), grep("3", names(CATS_wide))] = 0
pred2[c("prev_dep.7", "prev_sdq.7", "numeracy_score.7"), "numeracy_scoreW1"] = 0
```

A related approach is "two-fold FCS" proposed by Nevalainen et al. (2009). Two-fold FCS, uses a similar imputation model to FCS-1L-wide-MTW but imputations are carrried out using a two-step iterative process: iterations that cycle within a time-window and then iterations that cycle between time-windows.[42] It has been questioned whether the additional (within) iterative step is beneficial over the MTW approach,[21] and the two-fold approach is only available in Stata. Given these limitations, we do not discuss this approach further, although we provide code in GitHub for completeness.

## 3.2 MI approaches for imputing in long format

### 3.2.1 LMM-based JM (JM-2L)

This approach uses a multivariate LMM (MLMM), with cluster-specific random effects to model the correlations between repeated measurements of the same variable. As with JM-1L-wide, partially observed variables are included as outcomes while fully observed variables are included as predictors, with categorical variables imputed as continuous and potentially rounded in analysis; or handled using the latent formulation.

The JM-2L approach can be implemented in **jomo**. If the latent formulation is required, the variable needs to be stored as the correct type (continuous – "numeric"; categorical/binary – "factor") as for JM-1L-wide.

As with JM-1L-wide, the next step is to create data frames for the observed and incomplete variables, but with this approach separate data frames need to be specified for level 1 and level 2 variables as shown next.

We first create separate level 1 and level 2 data frames for the incomplete variables:

```r
# Level 1 variables
dataL_inc1 <- data.frame(prev_dep = CATS_long[, c("prev_dep")], numeracy_score = CATS_long[
   ,c("numeracy_score")])

# Level 2 variables
dataL_inc2 <- data.frame(ses = CATS_long[, c("ses")], numeracy_scoreW1 = CATS_long[,
   c("numeracy_scoreW1")])
```

Next, we create separate level 1 and level 2 data frames for the complete variables, which will be used as predictors with fixed effects (i.e. regression coefficients are assumed to be constant across individuals) in the imputation model:

```r
# Level 1 variables
dataL_compFE1 <- cbind(Intercept = rep(1, nrow(CATS_long)), CATS_long[,
   substr(names(CATS_long), 1, 6) %in% c("age", "sex", "prev_s", "time")])

# Level 2 variables
dataL_compFE2 <- cbind(Intercept = rep(1, nrow(CATS_long)), CATS_long[,
   substr(names(CATS_long), 1, 6) %in% c("age", "sex")])
```

With this approach, a data frame for complete variables with random (cluster-specific) effects also needs to be specified. In its simplest form, this could just be a vector of constants, to include only random intercepts for each incomplete variable, but can be extended to enable more general variance-covariance structures by including other time-varying complete variables for which we want to specify random effects or slopes (i.e., regression coefficients that vary across individuals). In the case study, we include a random slope for the *time* variable to make the imputation model more general than the analysis model:

```r
dataL_compRE <- cbind(Intercept = rep(1, nrow(CATS_long)), time = CATS_long[, "time"])
```

Imputations for JM-2L are then generated using the `jomo()` function as with JM-1L-wide, but the cluster indicator must be specified using the option `clus`: [43]

```r
set.seed(7251)
imp4 <- jomo(Y = dataL_inc1, Y2 = dataL_inc2, X = dataL_compFE1, X2 = dataL_compFE2,
   Z = dataL_compRE, clus = CATS_long$id, nimp = 66, nburn = 1000, nbetween = 100)
```

In the above specification we assumed constant covariance matrices for the error terms across clusters (individuals), which is the default option. It is also possible to allow for heterogeneous covariance matrices across individuals by assuming the covariance matrices follow a distribution, e.g., an inverse-Wishart distribution.[44] This can be achieved using the option `meth=random` in **jomo**. Assuming cluster-specific covariance matrices can be useful in settings where there is reason to believe that the clusters are heterogenous. [45] Some studies argue that imputation models with heterogenous covariance matrices can yield more reliable imputations than those with homogeneous covariance matrices due to improved compatibility of the conditional models.[46, 47] However, with small cluster sizes, assuming heterogeneous covariances can lead to overfitting of the cluster-specific parameters.[22, 46] In the context of longitudinal data, allowing for heterogeneity may not be very practical as often only a few repeated measurements are collected on each individual.[21] Nevertheless, if the substantive analysis model contains random slopes and the data allow, assuming heterogeneous individual-specific covariances can result in better estimation of subject-specific associations, although substantive model compatible JM approach is a more sound and theoretically justified way of handling same issue (c.f. Section 6: Discussion).[48] In CATS case study, we assume a homogenous covariance matrix across clusters due to the small number of repeated measures for each individual.

*3.2.2 LMM-based FCS (FCS-2L)*

The FCS-2L approach imputes missing values using a series of conditional LMMs specified for each incomplete variable, tailored according to the distribution of the variable. For imputing continuous incomplete time-varying variables, a LMM can be used while for categorical time-varying variables, either a LMM with latent variables or a generalized LMM (GLMM) with an appropriate link function can be used.[12, 49] The LMMs used to impute the repeated measures model the correlations between repeated measures of the same variable via individual-specific random effects.[20] Time-fixed variables are imputed by specifying the imputation model at the cluster (i.e. individual) level, with other time-fixed variables and aggregates (e.g., the mean) of the time-varying variables as predictors.[3, 20, 50]

The **mice**, **micemd** and **miceadds** R packages contain functions for implementing FCS-2L, identified by names starting with "2l.", and the user can choose an appropriate function based on the univariate imputation models required, as explained next. Note that there is an overlap in functionality of some functions from different packages (e.g., `2l.continuous` from **miceadds** is similar to `2l.lmer` from **mice**, in that they both specify a general LMM imputation model). A complete description of all available functions, including the imputation models specified in each for imputing incomplete continuous, binary and count variables, can be found in [20].

As with FCS-1L-wide, the variables first need to be stored as correct type before generating the imputations, as described earlier. The cluster variable must be of type integer:

```
CATS_dataL$c_id <- as.integer(CATS_dataL$c_id)
```

In the case study, the continuous time-varying variable `numeracy_score` can be imputed via `2l.pan` within **mice,** which fits a LMM with homogeneous within-cluster variances.[20] Alternatively, it is possible to fit a LMM with subject-specific residual error variances to allow additional flexibility; this can be achieved using `2l.norm`. We do not use this approach in the case study as it can be computationally intensive and extremely time-consuming. For the binary time-varying variable, `prev_dep,` we use a LMM imputation model with the latent variable formulation, implemented using `2l.jomo` (in **micemd**), as GLMM-based FCS approaches (using the logit link) often result in convergence issues (c.f. section 3.3).[51]

For continuous time-fixed variables such as `numeracy_scoreW1`, imputation can be carried out using `2lonly.norm` (in **mice**), which uses a linear regression model with aggregated mean values of the time-varying variables. For categorical time-fixed variable `ses` we use PMM, which samples imputations from the observed values of the same variable after fitting a regression model based in a nearest neighbour approach.[26] This can be implemented using the method 2lonly.pmm (in **mice**). See [20] for more details.

Setting up the imputation models for the CATS example as described is achieved using the code:

```
meth3 <- make.method(CATS_long)
meth3["prev_dep"] = "2l.jomo"
meth3["numeracy_score"] = "2l.pan"
meth3["ses"] = "2lonly.pmm"
```

```
meth3["numeracy_scoreW1"] = "2lonly.norm"
```

As with FCS-1L-wide, the next step is to set up the predictor matrix (named `pred3` below) where rows indicate the variable being imputed and the columns indicate the predictors used. Within this matrix, it is necessary to indicate the cluster group variable in each univariate imputation model. In **mice** this is achieved by setting the column corresponding to the cluster group (`id`) to "-2". In CATS example, as we are not yet accounting for the school level clustering (c.f. Section 4), we set the school ID variable to "0" in each row to omit it from the predictor sets for all variables. To specify random slopes for a predictor in a univariate imputation model for a certain variable, number "2" should be included in the corresponding column variable for the row variable that is being imputed. In the case study, we include random slopes for wave (*time*) in the imputation models for the time-varying variables.

As before, we use "1" to specify the predictors with fixed effects in the imputation models. These could be time-fixed predictors or time-varying predictors. For time-varying variables, the value of the predictor at a given timepoint is used to impute the time-varying incomplete variable at that time point. An alternative approach that has been discussed is to also include the cluster means of the time-varying predictors, in which case the number "3" should be included in the corresponding column variables. This approach has been argued to improve flexibility however currently there is scant evidence around this; some research has indicated that it has little impact in practice, while others suggest that including cluster means of incomplete (and even of complete variables) can be beneficial as these cluster means can serve as useful auxiliary variables.[20, 43] A pragmatic recommendation is to include the cluster means if possible given the sample size and the complexity of the imputation model. In the case study we include cluster means of all time-varying (complete and incomplete) predictors in the imputation models for all time-varying incomplete variables. The predictor matrix for the case study is set using the following code:

```
pred3[, c("school", "id")] = 0
pred3[, c("id")] = -2
pred3[c("ses", "numeracy_scoreW1"), c("time")] = 0
pred3[c("prev_dep", "numeracy_score"), c("time")] = 2
pred3[c("prev_dep"), c("numeracy_score", "prev_sdq")] = 3
pred3[c("numeracy_score"), c("prev_dep", "prev_sdq")] = 3
```

The imputations are then performed using the code:
```
imp5 = mice(data = CATS_long, m = 66, predictorMatrix = pred3, method = meth3, maxit = 10)
```

Finally, the analysis can be conducted as for *FCS-1L-wide*, except no reshaping is necessary prior to analysis.

### 3.3 Comparative performance of approaches for imputing longitudinal data

Imputing repeated measures in wide format allows an unstructured correlation among the incomplete repeated measures. It also means that no explicit distinction is made between repeated measures of the same and different variables or between time-fixed and time-varying variables. This approach is theoretically very flexible and is compatible with a range of substantive analysis models. It has been shown to provide reliable estimates of the regression coefficients in previous simulation studies with incomplete continuous variables when the substantive analysis is a random intercept model, random intercept model with interactions between incomplete variables and time and a logistic regression model fitted using generalised estimating equations (GEEs).[21, 41, 52] Imputing repeated measures in wide format has also been shown to perform well in the setting of a random intercept and random time-slope model with incomplete continuous variables.[48] Although theoretically mis-specified for categorical variables, simulations indicate that the standard JM approach to imputing in wide format (without the latent normal formulation) provides reliable estimates of all model parameters in the context of a random intercept model.[21] The standard FCS approach has also been shown to provide reliable estimates of all model parameters with incomplete binary, categorical and count data in the context of a random intercept model.[21] However it should be noted that the use of Poisson regressions within FCS for imputing count variables has been shown to induce bias in the regression coefficients of incomplete count variables.[22]

The standard JM and FCS approaches imputing in wide format have shown superior performance to approaches with moving time windows in simulations.[21, 41] However, these approaches can face convergence problems when (i) the proportion of missing data is high, (ii) there are many waves of data collection, (iii) there are large numbers of variables collected at each wave, especially those that are categorical, and (iv) the repeated measures are highly correlated (collinear).[41, 53, 54] If the standard approach does face convergence issues, a moving time window approach can be used, in which case it is preferable to use a wider time-window (e.g., 2 adjacent time points instead of only 1 adjacent time point).[55]

A major drawback of the approaches that impute data in wide format is that the repeated measures need to be balanced i.e. occur at aligned time-points across individuals. In cases with unbalanced longitudinal data, LMM-based MI approaches imputing in long format will be necessary.

LMM-based approaches explicitly distinguish time-varying variables from time-fixed variables. Correlations between the repeated measurements of the same variable are modelled via random effects assumed to follow a specific distribution, which means that LMM-based approaches are more restrictive compared to the approaches imputing in wide format. Although this can mean that the imputation model is less likely to face convergence problems, it requires the adoption of parametric assumptions about the random effects, which brings increased risk of bias if these assumptions are wrong.

Analytic and simulation findings have shown that both JM and FCS LMM-based approaches result in valid estimates of all model parameters in the context of a random intercept analysis model with incomplete continuous variables.[21, 43, 56] Similarly, in the context of a substantive analysis model with a random intercept and random time-slope with continuous incomplete variables, LMM-based approaches produce approximately unbiased estimates of the regression coefficients and variance components if only covariates are incomplete.[48] However, when both the outcome and covariates are incomplete, the LMM-based approaches result in biased estimates of the variance components due to incompatibility between the imputation and analysis model.[48, 57] In this context and where the longitudinal data are unbalanced, alternative MI approaches described in Section 5 are recommended.

The comparative performance of the various FCS and JM LMM-based MI approaches has also been investigated for handling incomplete binary, categorial and count variables via simulation, although all in the context of a random intercept only model. The findings suggest that with incomplete binary and count variables, GLMM-based FCS approaches can lead to biased estimates of variance components and regression coefficient estimates with confidence intervals exhibiting undercoverage[21, 22]. Furthermore, in GLMM-based FCS approaches, the logit link for binary and the log link for count variables can often cause singularity issues in estimating the covariance matrix, resulting in convergence issues.[19, 22, 51] In contrast, the LMM-based JM and FCS approaches that use the latent normal formulation result in approximately unbiased estimates of the regression coefficients and standard errors.[21, 46] However, it has been suggested that the JM approach can overestimate the variance components.[45, 46] The use of heterogeneous covariance matrices within JM and FCS approaches has also been shown to result in biased estimates of incomplete binary and count variables.[21, 22] Of note, both the JM and FCS approach extended with the latent variable formulation can be computationally intensive.[21, 22]

In summary, there is no single method that is universally preferable for imputing longitudinal data and the choice for best method depends on several factors including whether the data are balanced, which variable(s) are missing, which types of variables are missing, the substantive analysis model and the computational time. It should also be noted that, although we summarise some of the available guidance in the context of a few substantive analysis models above, the evaluation of MI methods for handling longitudinal data is still a developing area of research. A summary of current knowledge and available guidance for the methods are provided in Table 3.

**Insert Tables 3 and 4

### 4. MI approaches for longitudinal data with additional clustering

Some longitudinal studies include additional higher-level clustering due to naturally occurring clusters in the population. For example, in CATS individuals are clustered within schools. In such cases, the substantive analysis model would ideally account for both sources of correlation. In this section, we consider as substantive model a LMM similar to (1), but with random intercepts for both the school and the individual, as shown below:

$$\text{numeracy.score}_{ijk} = \beta_0 + \beta_1 * \text{prev.dep}_{ijk} + \beta_2 * k + \beta_3 * \text{numeracy.score}_{ij1} + \beta_4 * \text{sex}_j + \sum_{a=1}^{4} \beta_{5,a} * I[\text{SES}_{ij1} = a] + \beta_6 * \text{age}_{ij1} + b_{0i} + b_{0ij} + \varepsilon_{ijk} \quad (2)$$

where $i$ denotes the $i^{th}$ school, $j$ denotes the $j^{th}$ individual ($j = 1, \ldots 1200$), $k$ denotes the $k^{th}$ wave ($k = 3,5,7$), $\varepsilon_{ijk}$ denotes independent random measurement errors distributed as $\varepsilon_{ijk} \sim N(0, \sigma_1^2)$, $b_{0ij}$ denotes individual-level random effects distributed as $b_{0ij} \sim N(0, \sigma_2^2)$, and $b_{0i}$ denotes the school-level random effects distributed as $b_{0i} \sim N(0, \sigma_3^2)$. The rest of the notation is as described previously with the additional $i$ subscript. For simplicity we assume that individuals remain in the same schools at all three waves.

Within-cluster correlations arising due to higher-level clustering can be incorporated in the imputation model in two ways. They can be accommodated using random effects which imposes parametric assumptions on the correlation structure between clusters. Alternatively, if the number of clusters is relatively small, a set of indicator variables could be created to reflect the clustered structure and included in the imputation model. This is known as the dummy indicator (DI) approach. The inclusion of these indicators means a separate intercept (or a fixed effect) is modelled for each cluster allowing unrestricted variation between cluster means. If the substantive model includes incomplete variables with random slopes, to ensure compatibility it is required to include interactions between the indicator variables and other complete/incomplete variables as implied by the analysis model, which can mean estimating a lot of parameters.[20, 56] This can lead to non-convergence of the imputation models particularly when there is sparse data and in some cases can result in biased estimates of model parameters including variance components and higher-level regression coefficients (also cf. section 4.3). Note, in the previous section we did not consider the DI approach for modelling correlations arising due to longitudinal data (where a set of DIs is used to represent each individual in the sample) because in practice this is only feasible if there are few individuals and/or many observations per individual, which is rare.

We describe extensions of the approaches presented in section 3 to account for higher-level clustering in the following subsections.

### 4.1 MI approaches imputing in wide format

When there is higher-level clustering of longitudinal data, the repeated measures can be imputed in wide format with DIs or random effects to allow for the correlation due to higher-level clustering.

*4.1.1 Standard JM with dummy indicators for clusters (JM-1L-DI-wide)*

The JM-1L-wide approach can be applied with the DI approach to model between cluster correlations at the higher level. The approach can be implemented using **jomo** as before with JM-1L-wide but here the user will need to create DIs to reflect the higher-level clusters (schools in CATS) and include them in the data frame with other complete variables that are to be used as predictors in the imputation model. For the case study, this is achieved using the following code:

```
dataw_comp <- cbind(Intercept = rep(1, nrow(CATS_wide)), CATS_wide[, substr(names(CATS_wide),
    1, 6) %in% c("age", "sex", "prev_s", "school")])

dataw_comp <- dummy_cols(dataw_comp, select_columns = "school", remove_first_dummy = TRUE,
    remove_selected_columns = TRUE)
```

Fitting the imputation model, assessing convergence and extraction of imputed datasets can then be carried out as with JM-1L-wide. To fit the substantive model of interest on each imputed dataset, the original school cluster variable needs to be reattached to each of the imputed datasets (or recreated using the DIs) and each dataset then needs to be converted to long format for analysis. This can be achieved as shown below:

```
imp.long <- lapply(imp.list$imputations, function(d) {
    d$school <- CATS_wide$school
    reshape(d, varying = list(c("prev_dep.3", "prev_dep.5", "prev_dep.7"), c("numeracy_score.3",
        "numeracy_score.5", "numeracy_score.7"), c("prev_sdq.3", "prev_sdq.5", "prev_sdq.7")),
        idvar = "id", v.names = c("prev_dep", "numeracy_score", "prev_sdq"), times = c(3,
            5, 7), direction = "long")
})
```

The analysis model is then fitted to each imputed dataset using the code below, and results pooled as shown for JM-1L-wide. In CATS case study, convergence of the imputation model was achieved, but post-imputation analysis of one dataset generated non-convergence warnings. These warnings (in the `lme4()`) do not necessarily mean the fit is incorrect, and warnings can sometimes occur for apparently good fits. This was the case for our case study. Steps for resolution of these warnings can be found at [58].

```
fit.imp1 <- lapply(imp.long, function(d) {
    lmer(numeracy_score ~ prev_dep + time + age + numeracy_scoreW1 + sex + factor(ses) +
        (1 | school/id), data = d)
})
```

## 4.1.2 Standard FCS with dummy indicators for clusters (FCS-1L-DI-wide)

Similarly, the FCS-1L-wide approach can be used to model the repeated measures and extended to include DIs for clusters. This can be achieved via updating the predictor matrix to include the categorical cluster variable as a predictor in each of the univariate imputation models, with the remaining steps similar to *FCS-1L-wide* (cf. section 3.1.2). For CATS case study, the predictor matrix is specified using:

```
pred1 <- make.predictorMatrix(CATS_wide)
pred1[, c("id")] = 0
```

Similar to JM-1L-DI-wide, in CATS analysis post-imputation analysis generated non-convergence warnings in the analysis of one imputed dataset, which was a false positive.

## 4.1.3 LMM-based JM approach for clusters and imputing repeated measures in wide format (JM-2L-wide)

An alternative approach when there are repeated measures and higher-level clustering is to use the LMM-based JM approach (i.e. JM-2L) to model the cluster correlations via cluster-specific random effects, while still imputing the repeated measures in wide format. As with the JM-1L-wide approach, when implementing this approach using **jomo**, the first step is to store the incomplete variables as the correct variable type (binary and categorical incomplete variables as type "factor", and continuous variables as "numeric"):

```
cat.vars <- c("ses", "prev_dep.3", "prev_dep.5", "prev_dep.7")
CATS_wide[cat.vars] = lapply(CATS_wide[cat.vars], factor)
```

Next, data frames for the observed and incomplete variables, and a data frame with random (cluster-specific) effects, needs to be specified. In the case study, we only specify a random intercept for the school clusters, which is achieved by including a column of 1's in the random effects data frame using the code:

```
#Data frame(s) with variables to be imputed

  #Level 1 variables (individual-specific variables)
  dataw_inc <- CATS_wide[, substr(names(CATS_wide), 1, 6) %in% c("prev_d", "numera", "ses")]

  #Need to create a separate level 2 variable data frame if there are incomplete cluster
  (school) specific variables

#Data frame(s) with complete variables to be used as predictors of the imputation model

  # Level 1 variables
  dataw_compFE <- cbind(Intercept = rep(1, nrow(CATS_wide)), CATS_wide[,
    substr(names(CATS_wide), 1, 6) %in% c("age", "sex", "prev_s")])

  #Need to create a separate level 2 variable data frame if there were incomplete cluster
  (school) specific variables

#Data frame with complete variables to be used as predictors of the imputation model, with
RANDOM EFFECTS
  dataw_compRE <- cbind(Intercept = rep(1, nrow(CATS_wide)))
```

The data frame with cluster-specific random effects (i.e. `dataw_compRE` above) can be extended to enable more general variance-covariance structures. This is achieved by including other complete variables in this data frame to allow random effects (i.e. random slopes) for these variables.

Next, the imputations are run using the code below with the cluster indicator specified using the option `clus`:

```
imp3 <- jomo(Y = dataw_inc, X = dataw_compFE, Z = dataw_compRE, clus = CATS_wide$school,
    meth = "random", nimp = 66, nburn = 1000, nbetween = 1000)
```

As with JM-1L-wide, post imputation, each dataset needs to be converted to long format with the substantive analysis model fitted to each, followed by pooling of results.

### 4.1.4 LMM-based FCS approach for clusters while imputing repeated measures in wide format (FCS-2L-wide)

A similar extension to JM-2L can also be applied to FCS-2L, with LMM-based univariate imputation models to model the cluster correlations via cluster-specific random effects, with the repeated measures imputed in wide format. As with FCS-2L, if there are incomplete cluster-level variables, they are imputed by specifying an imputation model at the cluster level with all other cluster-level variables (if there are any) and aggregates of all time-varying and time-fixed variables as predictors.

As with FCS-2L, the variables need to be stored as correct type with the cluster variable stored as integers:

```
Set categorical and binary variables to impute as factors
cat.vars <- c("ses", "prev_dep.3", "prev_dep.5", "prev_dep.7")
CATS_wide[cat.vars] = lapply(CATS_wide[cat.vars], factor)

Set cluster variable to integer
CATS_wide$school <- as.integer(CATS_wide$school)
```

Similar to FCS-2L, we impute the continuous variables (time-varying and time-fixed) using LMMs with homogenous within-group variances (via `2l.pan` in **mice**), binary variables using a LMM with the latent variable formulation (via `2l.jomo` in **micemd**), and using PMM for categorical variables (via `2l.pmm` in **miceadds**):

```
meth2 <- make.method(CATS_wide)
meth2[substr(names(CATS_wide), 1, 6) %in% c("prev_d")] = "2l.jomo"
meth2[substr(names(CATS_wide), 1, 6) %in% c("numera")] = "2l.pan"
meth2[substr(names(CATS_wide), 1, 5) %in% c("ses")] = "2l.pmm"
```

Next, as before with FCS-2L, the predictor matrix needs to be specified and updated to indicate the cluster variable in each univariate imputation model by setting the column corresponding to the cluster variable to "-2". Because we are accounting for the individual level clustering by imputing in wide format with this approach, we set the individual ID variable to "0" in each column to omit this variable from univariate imputation models. This is achieved using the following:

```
pred2 <- make.predictorMatrix(CATS_wide)
pred2[, c("id")] = 0
pred2[, c("school")] = -2
```

Finally, imputations are generated using the following code:

```
imp4 = mice(data = CATS_wide, m = 66, predictorMatrix = pred2, method = meth2, maxit = 10)
```

Post imputation, each dataset needs to be converted to long format with the substantive analysis model fitted to each, followed by pooling of results (cf. section 3.1.1). In the case study, non-convergence warnings were generated for one of the imputed datasets but given the small proportion we ignore these.

## 4.2 MI approaches imputing in long format

As with longitudinal studies where there is no higher-level clustering, as alternative method for incorporating the repeated measures in the imputation model is to impute them using a LMM-based approach applied to the data in long format as described in section 3.2. As with the wide approaches, the long format approaches can be extended to allow for higher level clustering using either a DI approach or using random effects.

### *4.2.1 LMM-based JM approach imputing in long format with DI for school clusters (JM-2L-DI)*

The JM-2L approach applied to repeated measures in long format can be extended using DIs to model the higher-level clustering.

The approach can be implemented in **jomo,** with similar steps as JM-2L**,** except that here we will need to create DIs to reflect the higher-level clusters, which need to be included in the data frames at both level 1 and level 2 along with other complete variables that are to be used as predictors. This can be achieved as follows:

```
Data frame with complete variables to be used as predictors of the imputation model, with only
FIXED EFFECTS

  # Level 1 variables
  dataL_compFE1 <- cbind(Intercept = rep(1, nrow(CATS_long)), CATS_long[,
    substr(names(CATS_long), 1, 6) %in% c("age", "sex", "prev_s", "time", "school")])

  #Create dummy indicators for school cluster variable
  dataL_compFE1 <- dummy_cols(dataL_compFE1, select_columns = "school",
    remove_first_dummy = TRUE, remove_selected_columns = TRUE)

  # Level 2 variables
  dataL_compFE2 <- cbind(Intercept = rep(1, nrow(CATS_long)), CATS_long[,
   substr(names(CATS_long), 1, 6) %in% c("age", "sex", "school")])

  #Create dummy indicators for school cluster variable
  dataL_compFE2 <- dummy_cols(dataL_compFE2, select_columns = "school",
    remove_first_dummy = TRUE, remove_selected_columns = TRUE)
```

Imputations are then generated using the following code with other steps similar to JM-2L:

```
imp5 <- jomo(Y = dataL_inc1, Y2 = dataL_inc2, X = dataL_compFE1, X2 = dataL_compFE2,
    Z = dataL_compRE, clus = CATS_long$c_id, meth = "random", nimp = 66, nburn = 1000,
    nbetween = 1000)
```

Note that as with JM-1L-DI-wide, here post imputation analysis generated warnings in one of the imputed datasets but given the small proportion we ignore these.

### 4.2.2 LMM-based FCS approach imputing in long format with DI for school clusters (FCS-2L-DI)

Similarly, the FCS-2L approach can be extended by including DIs for clusters as predictors in the univariate LMM-based imputation models.

In our experience, FCS-2L-DI often faces convergence issues in the imputation model due to sparse data, limiting its use in practice [23, 59] This was observed in the CATS case study. Given these convergence issues we omit the illustration of this approach, although code for implementing it is available in the GitHub repository.

### 4.2.3 LMM-based JM approach imputing in long format (JM-3L)

Rather than using a LMM to model the repeated measures and DIs to model the higher-level clusters, the LMM approach can be extended to include a second random effect to reflect the higher-level clusters. This can be achieved by extending the JM-2L approach to include cluster-specific random effects. Implementations of this approach are available in Stat-JR and MPlus, [60, 61] but not in R or Stata. Therefore, we do not discuss this approach further in this paper.

### 4.2.4 LMM-based FCS approach imputing in long format (FCS-3L)

The FCS-2L approach can also be extended to include cluster-specific random effects to allow for the higher-level clustering.[20]

The FCS-3L approach can be implemented using the **miceadds** package in R via `ml.lmer()`. Note the syntax used for specifying the imputation model in `ml.lmer()` is somewhat different from the other univariate functions within **mice**, as illustrated next.

First, in contrast to previous specifications of FCS approaches, all categorical and binary variables need to be stored as numeric variables coded as 0,1,2,…:

```
cat.vars <- c("ses", "prev_dep")
CATS_long[cat.vars] = lapply(CATS_long[cat.vars], as.numeric)
CATS_long$ses <- CATS_long$ses - 1
CATS_long$prev_dep <- CATS_long$prev_dep - 1
```

Second, the clustered structure of the variables is not specified using the predictor matrix as with previous FCS approaches, but instead is specified by:

i. Specifying the level at which each incomplete variable is measured, i.e. whether they are measured at individual level or at school level. In order to do this, we first create a vector to store the levels. In the case study, as we only have missing values in time-fixed variables measured at the individual level (level 2) and time-varying variables (level 1), we specify that for the incomplete time-fixed variables the cluster level is at the individual level ("`c_id`"), and the lowest level variables are left blank.

```
variables_levels <- miceadds:::mice_imputation_create_type_vector(colnames(CATS_long)
    ,value = "")
variables_levels["ses"] <- "id"
variables_levels["numeracy_scoreW1"] <- "id"
```

ii. Specifying the hierarchical structure of the incomplete variables: This is achieved by specifying within which clusters each incomplete variable is clustered (for e.g., time-varying variables are clustered within individual and school clusters, time-fixed variables measured for each individual are clustered within school clusters) in a list as shown below:

```r
cluster <- list()
cluster[["prev_dep"]] <- c("id", "school")
cluster[["numeracy_score"]] <- c("id", "school")
cluster[["ses"]] <- c("school")
cluster[["numeracy_scoreW1"]] <- c("school")
```

The third step is to specify the univariate imputation methods to be used in the imputation procedure. For time-varying variables and time-fixed variables we use the `ml.lmer()` in **miceadds.** The `ml.lmer()` function can handle continuous (using a LMM), binary (using a GLMM with a logit link) and ordinal data (using PMM), and the univariate imputation model for each variable should be specified as shown next. For the case study, we use PMM for categorical variables and a general LMM for continuous variables:

```r
# method for imputing lower-level variables (time-varying and time-fixed)
  meth4 <- mice::make.method(data = CATS_long)
  meth4[c("prev_dep", "numeracy_score", "ses", "numeracy_scoreW1")] <- "ml.lmer"

# Specify univariate imputation models(impute c_ses and prev_dep with pmm and others with LMMs)
  model <- list(prev_dep = "pmm", ses = "pmm", numeracy_score = "continuous",
    numeracy_scoreW1 = "continuous")
```

Note that the `ml.lmer()` function is only required for incomplete time-varying (i.e. level 1) variables where there are two sources of clustering; for incomplete variables at higher levels, users may use other univariate imputation models discussed under *FCS-2L*.

Next the predictor matrix needs to be specified. Note below that in contrast to other univariate methods in mice, for variables imputed with `ml.lmer()` cluster indicators are not included. This is because for these variables, hierarchical structure was specified earlier

```r
pred4 <- mice::make.predictorMatrix(data = CATS_long)
pred4[, c("id", "school")] <- 0
pred4[("numeracy_scoreW1"), c("time")] <- 0
pred4[("ses"), c("time")] <- 0
```

Finally imputations are performed using the following code, where the hierarchical structure is specified using the option `levels_id,` and the level of each incomplete variable is specified using the option `variable_levels` with the imputation models for those variables using `ml.lmer()`specified using the option `model`:

```r
imp7 <- mice(CATS_long, method = meth4, predictorMatrix = pred4, maxit = 10, m = 66,
    levels_id = cluster, variables_levels = variables_levels, model = model)
```

Post-imputation, the analysis of interest is fitted to each imputed dataset using similar syntax as shown for FCS-1L-wide previously with no reshaping.

### 4.3 Comparative performance of approaches for imputing longitudinal data with additional clustering

A limited number of simulation studies have shown that all of the approaches described in this section provide valid results when the substantive analysis model is a random intercept only model, while the approaches imputing in wide format can provide valid results for a random intercept substantive model with an interaction between the time-varying exposure and time.[23, 52] This is promising as JM-1L-DI-wide and FCS-1L-DI-wide are the only approaches that can be implemented in software that do not currently support MI approaches for clustered data such as Stata. However, the approaches that use DIs should be used with caution as the DI approach can only provide valid results under limited conditions. While the DI approach has been shown to produce approximately unbiased estimates of regression coefficients in the context of a random intercept substantive analysis model when there are missing values in the outcome, and unbiased estimates of regression coefficients at level 1 when there are missing values in the outcome and covariates,[14, 56, 62, 63] they have subpar performance in other situations. First, the DI approach overestimates the standard errors of regression coefficients and the variance components, when the intra cluster correlation (ICC) is low, or the cluster sizes are small.[56, 62] Second, the DI approach overestimates higher-level variance components when there is missing outcome data,[56, 62] and results in biased estimates of the higher-level regression coefficients when there is missing exposure data.[56, 62] Third, when data are missing completely in one or more clusters, fixed effects corresponding to the DIs of these clusters can face identifiability issues in the imputation model. Finally, when using the DI approach, it is not possible to include variables that are constant within the cluster (e.g., school-level variables in the CATS), as these variables are collinear with cluster membership.[62] In all of these cases LMM-based approaches should be used, with LMM-based approaches imputing repeated measures in long format being the only option if the repeated measures are unbalanced.[64]

A summary of current knowledge and available guidance for the methods are provided in Table 5.

**Insert Tables 4 and 5

### 5. Considerations for choosing a method in practice: Illustration with CATS case study

In this section, we describe the process for choosing an MI approach among those reviewed in sections 3 and 4 using the CATS case study as an illustration. We start with the assumption that MI is the most appropriate approach to handle missing data. The decision making process on how best to handle missing data is explained in detail elsewhere.[65]

**5.1 Longitudinal data with no additional clustering**

If the substantive analysis model only includes a random intercept, as in substantive analysis model (1) from the CATS case study, any of the approaches outlined in section 3, with the exception of FCS-1L-wide-MTW, should in theory provide reliable estimates (see Table 2). However, given the differences between computational time in R (see supplementary file: section S5), we would choose FCS-1L-wide which is the simplest approach, and has the least computational time among all approaches. To achieve compatibility with (1), the imputation model specification should proceed as described in section 3.1.2, with all analysis variables in the same form as in the analysis and auxiliary variables (if any), with the number of burn in iterations and number of imputations set as illustrated. Although with FCS a low number of iterations are usually enough to achieve convergence, we recommend examining the trace plots to assess convergence and the number of iterations be decided on a case-by-case basis.

Similarly, in Stata we would choose FCS-1L-wide given its simplicity and to its ability to handle categorical incomplete variables within the imputation process without the need for post-rounding approaches that can introduce bias. [31]

**5.2 Longitudinal data with additional level of clustering**

If there were longitudinal data with higher level clustering and the substantive analysis model of interest was in the form of equation (2) from CATS, any of the approaches in section 4 would be appropriate (see Table 4). However, given the well-documented issues of the DI approach in some settings,[23, 62] it is advisable to use approaches that do not use DIs for the higher-level clustering. Among the approaches that do not use DIs s, in R we would opt for JM-2L-wide for a number of reasons: (1) no post-imputation analysis warnings were generated, (2) the computational time is lower compared to other approaches, and (3) the latent normal formulation for binary and categorical variables allows handling incomplete binary and categorical variables flexibly within the imputation procedure. The imputation model specification would then proceeds as illustrated in section 4.1.3. When using this approach we would recommend using at least 500 burn-in iterations as JM approaches usually require a much larger number of burn-in iterations compared to FCS (e.g.500-1000 vs. 5-20) to achieve convergence. As with FCS, we recommend examining the trace plots to determine the appropriate number of burn in iterations. The default setting of between imputation iterations in **jomo** (1000) should be sufficient to ensure independence between imputed datasets in most cases. While the number of between imputations could be reduced to save computational time if needed, we recommend using at least 100.

If using Stata, the only available approaches are JM-1L-DI-wide and FCS-1L-DI-wide. Of these approaches we would recommend using FCS-1L-DI-wide due to its capability to handle binary and categorical variables within the imputation process.

6. **Discussion**

The variety of MI approaches that can be used for handling incomplete longitudinal data has grown over recent years, with several implementations available in mainstream software. However, these approaches can be challenging to implement due to their complexity. In this tutorial we illustrate the use of several of these MI approaches in the context of longitudinal and clustered longitudinal data using R (and Stata where applicable). We focus on an LMM substantive analysis model for simplicity, but there is a wide array of analysis approaches such GEEs or LMMs with more complex random effect structures, that can be used for analysing longitudinal data depending on the research question of interest. A detailed discussion of potential analysis approaches is beyond the scope of the current paper, but can be found elsewhere.[59, 66]. A discussion of how to impute incomplete data for different substantive models is also beyond the scope of this paper, but we summarize the available guidance in sections 3.3 and 4.3.

In this paper we focus on the standard implementations of JM and FCS approaches that are widely available in statistical software. An alternative and increasingly popular approach to ensure compatibility between the imputation and the analysis model is to use a substantive model compatible (SMC) MI approach.[11] SMC MI tailors the imputation model to the substantive analysis model of interest by specifying the imputation model as the product of the joint distribution of non-outcome variables and the conditional distribution for the outcome given non-outcome variables, with the latter aligning with the substantive analysis model.[11, 67] Specifying the imputation model in this way ensures compatibility between the imputation and analysis models. SMC MI approaches can be especially useful when the analysis model includes complex terms such as polynomial terms, interactions, or a random slope,[11, 57, 68, 69] where specifying compatible conditional models can be challenging.

The imputation process of an SMC MI approach is similar to that of the standard MI methods, with an additional Metropolis-Hastings step to accept or reject the draws to selectively sample from the desired distributions as implied by the imputation model.[70] It can be implemented in both the JM and FCS frameworks.[11, 70] The SMC MI approach has shown superior performance compared to standard MI approaches in simulation studies in longitudinal settings where the analysis model contains polynomial terms, interactions, or a random slope.[11, 48, 52, 71] While some implementations of SMC MI for longitudinal/clustered data can be found in **jomo**, **mdmb** and in a standalone software **Blimp**,[17, 72, 73] development and evaluation of SMC MI approaches that can cater to a wide array of substantive analysis models in mainstream software is still an ongoing area of research. Illustration of these approaches was beyond the scope of this manuscript.

We focused on R and Stata for illustrating the available MI approaches for imputing longitudinal data as they are widely used for epidemiological analyses, but there are other software that can be used to conduct MI for clustered data, including SAS, Mplus, REALCOM, Blimp and Stat-JR**.**[60, 73, 74] While applied researchers have many MI implementations to choose from for imputing longitudinal data, current implementations, especially those for binary/categorical data and three-level data, would benefit from having faster algorithms, being more robust to model breakdowns (i.e., termination of the imputation procedure without imputed values being generated), having meaningful warning/error messages in the case of model breakdowns and having wider availability of guidance. We hope that the illustrations and guidance on the MI methods in the current manuscript improve understanding and encourage their uptake among practitioners leading to greater compatibility of MI models with the analysis of longitudinal data, thereby enhancing the quality of results.

**Acknowledgements**
MQ and IRW were supported by the Medical Research Council Programme MC_UU_00004/07. KL was supported by an NHMRC investigator grant level 1, ID 2017498. MMB was supported by funding from the Australian National Health and Medical Research Council (NHMRC) Investigator Grant 2009572.

## Tables

Table 1: Description of CATS variables based on a simulated dataset of n=1200 individuals.

| Variable | Variable label | | Type of variable | Grouping / Range | N with Missing data (%) |
|---|---|---|---|---|---|
| | **Wide format** | **Long format** | | | |
| **Child's sex assigned at birth** | sex | sex | Categorical | 0 = Female, 1 = Male | 0 |
| **Child's age (wave 1)** | age | age | Continuous | Range [7-11] | 0 |
| **SES measured by the SEIFA IRSAD quintile (wave 1)** | ses | ses | Categorical | 0=1st quintile (most disadvantaged) | 269 (22.4%) |
| | | | | 1=2nd quintile | |
| | | | | 2=3rd quintile | |
| | | | | 3=4th quintile | |
| | | | | 4=5th quintile (most advantaged) | |
| **Standardized numeracy score (wave 1)** | numeracy_scoreW1 | numeracy_scoreW1 | Continuous | z-score | 192 (16.0%) |
| **Standardized numeracy score** | | | Continuous | z-score | |
| Wave 3 | numeracy_score.3 | numeracy_score | | | 187 (15.6%) |
| Wave 5 | numeracy_score.5 | | | | 293 (24.4%) |
| Wave 7 | numeracy_score.7 | | | | 419 (35.0%) |
| **Depressive symptoms** | | | Binary | 0 = No, 1 = Yes | |
| Wave 2 | prev_dep.3 | prev_dep | | | 154 (12.8%) |
| Wave 4 | prev_dep.5 | | | | 192 (16.0%) |
| Wave 6 | prev_dep.7 | | | | 268 (22.3%) |
| **Child behaviour reported by SDQ (auxiliary variable)** | | | Continuous | Range [0-40] | |
| Wave 2 | prev_sdq.3 | prev_sdq | | | 0 |
| Wave 4 | prev_sdq.5 | | | | 0 |
| Wave 6 | prev_sdq.7 | | | | 0 |

IRSAD: Index of Relative Socio-Economic Advantage and Disadvantage, NAPLAN: National Assessment Program - Literacy and Numeracy, SDQ: Strengths and Difficulties Questionnaire, SEIFA: Socioeconomic Index for Areas, SES: Socio-Economic Status.

Table 2: Summary of approaches for imputing longitudinal data: properties

| | **Method** | | | | |
|---|---|---|---|---|---|
| | *JM-1L-wide* | *FCS-1L-wide* | *FCS-1L-wide-MTW* | *JM-2L* | *FCS-2L* |
| Can the approach handle longitudinal data that do not occur at aligned time-points across individuals? | No | No | No | Yes | Yes |
| How are the correlations among repeated measures of the same variable handled? | If incomplete: via the unstructured covariance matrix of the MVN model; If complete: via regression; both allow an unrestricted correlation structure | Via regression, allows an unrestricted correlation structure | Via regression, allows an unrestricted correlation structure but assumes measures being imputed are conditionally independent of the variables outside the time window | Via random effects, forces a restricted correlation structure | Via random effects, forces a restricted correlation structure |
| How are the correlations among repeated measures of different variables handled? | If incomplete: via the unstructured covariance matrix of the MVN model If complete: via regression | Via regression | Via regression if within the specified time window, otherwise assumed conditionally independent | If incomplete: via the unstructured covariance matrix of the MVN model If complete: via regression | Via regression |
| Explicit distinction between time-varying and time-fixed variables? | No | No | No | Yes | Yes |
| Explicit distinction between repeated measures of the same and different variables? | No | No | No | Yes | Yes |

Table 3: Available guidance on approaches for imputing longitudinal data

| | Method | | | | |
|---|---|---|---|---|---|
| | *JM-1L-wide* | *FCS-1L-wide* | *FCS-1L-wide-MTW* | *JM-2L* | *FCS-2L* |
| Performance* (with continuous incomplete variables) when substantive model is | | | | | |
| (i) Random intercept only model | Provides reliable estimates of all model parameters | Provides reliable estimates of all model parameters | Results in biased estimates and under coverage of regression coefficient parameters | Provides reliable estimates of all model parameters[+] | Provides reliable estimates of all model parameters[+] |
| (ii) Random intercept and time-slope model | Provides reliable estimates of all model parameters | Provides reliable estimates of all model parameters | No clear guidance | Provides reliable estimates of all model parameters if only the covariates are incomplete.[+] When both the outcome and covariates are missing, provides biased variance components. | Provides reliable estimates of all model parameters if only the covariates are incomplete.[+] When both the outcome and covariates are missing, provides biased variance components. |
| (iii) Generalized estimating equations with an unstructured correlation | Provides reliable estimates of all model parameters | Provides reliable estimates of all model parameters | Provides reliable estimates of all model parameters but slightly biased and less precise compared to *JM-1L-wide* and *FCS-1L-wide* and very narrow time-windows have led to poor performance | No clear guidance | No clear guidance |
| Performance* (with categorical and count incomplete variables) when substantive model is | | | | | |
| (i) Random intercept only model | Provides reliable estimates of all model parameters | Provides reliable estimates of all model parameter. Can lead to biased estimates regression coefficients with Poisson regression for count variables. | Results in biased estimates of regression coefficient parameters and undercoverage for regression coefficient parameters. | Provides reliable estimates of the regression parameters$^{+}$, can overestimate variance component estimates. Use of heterogenous covariance matrices can result in biased estimates of incomplete binary and count variables. | Can lead to regression coefficient estimates with undercoverage and result in biased estimates of variance components. Use of heterogenous covariance matrices can result in biased estimates of incomplete binary and count variables. |
| (ii) Random intercept and time-slope model | No clear guidance | No clear guidance | No clear guidance | No clear guidance | No clear guidance |
| (iii) Generalized estimating equations with an unstructured correlation | No clear guidance | No clear guidance | No clear guidance | No clear guidance | No clear guidance |
| Convergence issues | | High risk, and can happen with<br>　i. high proportion of missing data<br>　ii. many waves of data collection<br>　iii. large numbers of variables at each wave (especially if categorical)<br>　iv. highly collinear repeated measures | Low risk with a sufficiently narrow time-window is specified | Low risk | Low risk |

*Performance is said to be reliable if method results in approximately unbiased estimates with nominal coverage

$^{+}$May result in biased results due to incorrect distributional assumptions about the random effects in the imputation model

Note: "No clear guidance" indicates that there is no clear guidance via simulations in the setting of longitudinal data. While some guidance is available for other clustered settings in some of these contexts, we exclude these as it is not clear if these findings can be extrapolated to longitudinal settings; Guidance provided assumes that the imputation model is otherwise appropriately tailored and that the substantive analysis is correctly specified

Table 4: Summary of approaches for imputing longitudinal data with additional level of clustering: properties

| | Method | | | | | | | |
|---|---|---|---|---|---|---|---|---|
| | *JM-1L-DI-wide* | *FCS-1L-DI-wide* | *JM-2L-wide* | *FCS-2L-wide* | *JM-2L-DI* | *FCS-2L-DI* | *JM-3L* | *FCS-3L* |
| Can the approach handle longitudinal data that do not occur at aligned time-points across individuals? | No | No | No | Yes | Yes | Yes | Yes | Yes |
| How are the correlations among individuals from the same higher-level cluster are handled? | Using DIs | Using DIs | Via random effects | Via random effects | Using DIs | Using DIs | Via random effects | Via random effects |
| How are the correlations among repeated measures of the same variable are handled? | If incomplete: via the unstructured covariance matrix of the MVN model, If complete: via regression; both allows an unrestricted correlation structure | Via regression, allows an unrestricted correlation structure | If incomplete: via the unstructured covariance matrix of the MVN model, If complete: via regression; both allows an unrestricted correlation structure | Via regression, allows an unrestricted correlation structure | Via random effects, forces a restricted correlation structure | Via random effects, forces a restricted correlation structure | Via random effects, forces a restricted correlation structure | Via random effects, forces a restricted correlation structure |
| How are the correlations among repeated measures of different variables are handled? | If incomplete: via the unstructured covariance matrix of the MVN model, If complete: via regression; allows an unrestricted correlation structure | Via regression, allows an unrestricted correlation structure | If incomplete: via the unstructured covariance matrix of the MVN model, If complete: via regression; allows an unrestricted correlation structure | Via regression, allows an unrestricted correlation structure | If incomplete: via the unstructured covariance matrix of the MVN model If complete: via regression | Via regression | If incomplete: via the unstructured covariance matrix of the MVN model If complete: via regression | Via regression |

| | | | | | | | |
|---|---|---|---|---|---|---|---|
| Explicit distinction between time-varying and time-fixed variables? | No | No | No | No | Yes | Yes | Yes | Yes |
| Explicit distinction between repeated measures of the same and different variables? | No | No | No | No | Yes | Yes | Yes | Yes |

Table 5: Available guidance on approaches for imputing longitudinal data with additional level of clustering

| | Method | | | | | | |
|---|---|---|---|---|---|---|---|
| | *JM-1L-DI-wide* | *FCS-1L-DI-wide* | *JM-2L-wide* | *FCS-2L-wide* | *JM-2L-DI* | *FCS-2L-DI* | *FCS-3L* |
| Performance* (with continuous incomplete variables) when substantive model is | | | | | | | |
| (i) Random intercept only model | Provides reliable estimates of the exposure- regression coefficient# | Provides reliable estimates of the exposure- regression coefficient# | Provides reliable estimates of the exposure- regression coefficient+ | Provides reliable estimates of the exposure- regression coefficient+ | Provides reliable estimates of the exposure-regression coefficient+# | Provides reliable estimates of the exposure-regression coefficient +# | Provides reliable estimates of the exposure-regression coefficient+ |
| (ii) Random intercept model with an interaction between time-varying exposure and time | Provides reliable estimates of the exposure- regression coefficient and the interaction term# | Provides reliable estimates of the exposure- regression coefficient and the interaction term # | Provides reliable estimates of the exposure- regression coefficient and the interaction term + | Provides reliable estimates of the exposure- regression coefficient and the interaction term + | Substantive model compatible adaptations (c.f. section 6) of these allow reliable estimation of the exposure- regression coefficient and the interaction term+ | | |
| (iii) Random intercept model with other general interactions and/or non-linear terms | Results in biased estimates of the exposure- regression coefficient and the interaction/nonlinear | Results in biased estimates of the exposure- regression coefficient and the interaction/nonlinear | Results in biased estimates of the exposure- regression coefficient and the interaction/nonlinear | Results in biased estimates of the exposure- regression coefficient and the interaction/nonlinear | Substantive model compatible adaptations (c.f. section 6) of these allow reliable estimation of the exposure- regression coefficient and the interaction/nonlinear term(s) + | | |

| | | | | |
|---|---|---|---|---|
| | term(s) | term(s) | term(s) [+] | term(s) [+] |
| Convergence issues | Very high risk, and can happen with<br>i. high proportion of missing data<br>ii. many waves of data collection and/or DIs<br>iii. large numbers of variables at each wave (especially if categorical)<br>iv. highly collinear repeated measures<br>v. data are missing completely in one or more clusters | High risk, and can happen with<br>i. high proportion of missing data<br>ii. many waves of data collection<br>iii. large numbers of variables at each wave (especially if categorical)<br>iv. highly collinear repeated measures | High risk, and can happen with<br>i. high proportion of missing data<br>ii. large numbers of DIs<br>iii. data are missing completely in one or more clusters | Low risk |

\* Performance is said to be reliable if method results in approximately unbiased estimates with nominal coverage

[+] May result in biased results due to incorrect distributional assumptions about the random effects in the imputation model

[#] Can overestimate standard errors and variance components when the intra cluster correlation is low, or the cluster sizes are small.

Note: JM-3L omitted due to unavailability of software; Guidance provided assumes that the imputation model is otherwise appropriately tailored and that the substantive analysis is correctly specified

**Supplementary files**

*S1: CATS case study analysis results based on available cases.*

Table S1: Point estimate (and standard error) for the effect of early depressive symptoms on subsequent standardized NAPLAN numeracy scores and point estimates for the variance components (VC) at levels 3, 2 and 1.

| Model | Regression coefficient (SE) | VC- school level | VC-individual level | Residual error |
|---|---|---|---|---|
| Equation (1) | 0.023 (0.038) | - | 0.278 | 0.253 |
| Equation (7) | 0.020 (0.037) | 0.044 | 0.238 | 0.252 |

*S2: Directed acyclic graph (DAG) for the substantive analysis.*

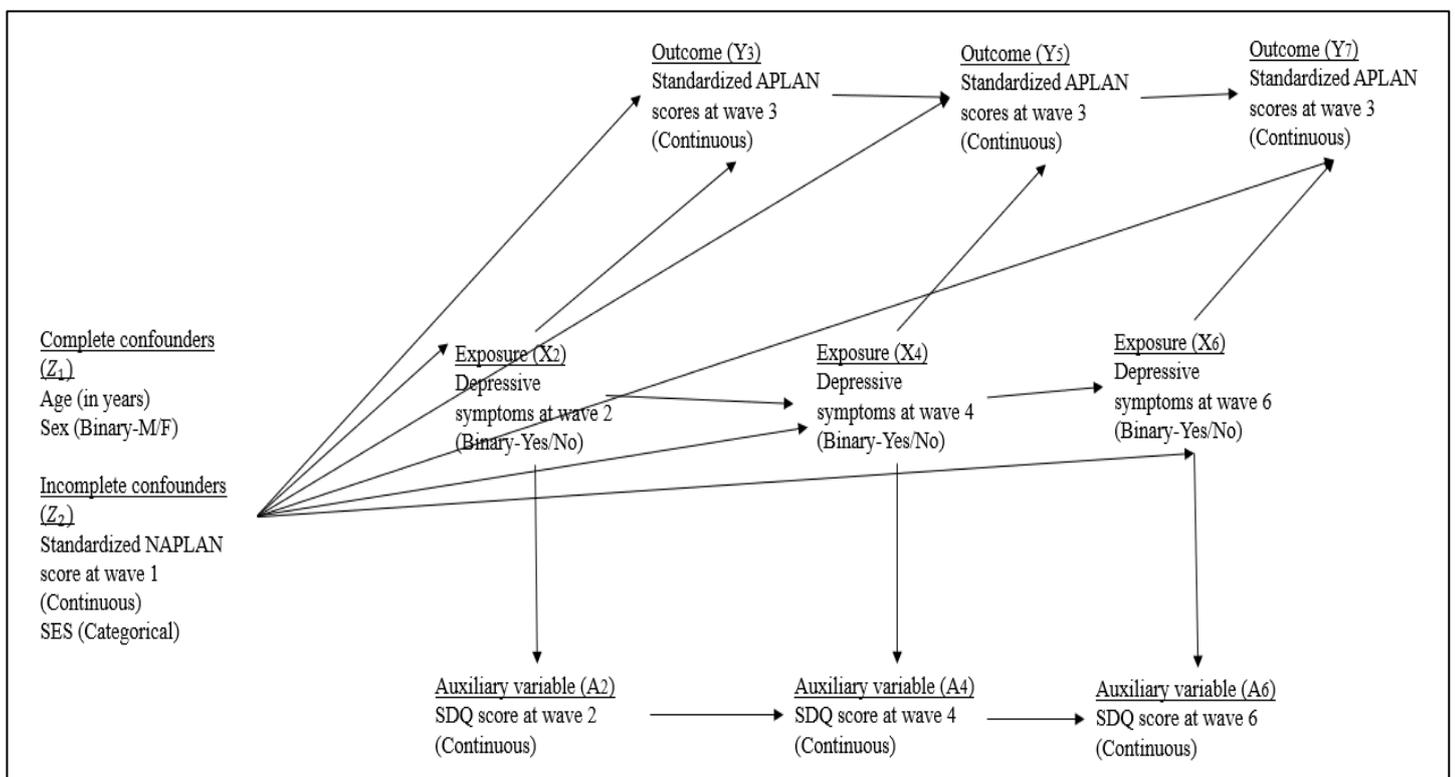

Figure A1: Directed acyclic graph illustrating the causal assumptions in the substantive analysis

*S3: Simulation of the dataset used in this tutorial*

Simulation of complete data

First we generated 40 school clusters which were populated to contain a varying number of students ranging from 8-66 students, similar to the CATS data [1]. The school cluster sizes ($8 \leq n_i \leq 66$) were assumed to follow a truncated log-normal distribution and cluster size for each school $i$ was sampled randomly from this distribution. In order to set the total number of students across the 40 schools to be 1200, the sampled cluster sizes were multiplied by a factor of $1200/\sum_{i=1}^{40} n_i$ and rounded to derive a scaled class size. If the total of these scaled class sizes was less than 1200, the deficit was added to the last school cluster, if the total of scaled class size was higher than 1200, the excess was deducted from the last school cluster.

The rest of the variables were then generated sequentially as described below for individual $j$ in cluster $i$. The values of the parameters indexing these distributions were determined by estimating the respective quantity from the CATS data.

i. Child's age at wave 1 ($age_{ij1}$) was generated from a uniform distribution, $U(7,10)$.

ii. Child's sex ($sex_{ij}$) was generated by randomly assigning 50% of students to be female.

iii. Child's SES quintile at wave 1 ($SES_{ij1}$) was generated by randomly assigning 10%,10%,20%, 30% and 30% of respondents to SES quintiles 1,2,3,4 and 5 respectively.

iv. The standardised NAPLAN scores at wave 1 ($Y_{ij1}$) were generated from a linear regression model conditional on child's sex, child's age at wave 1 and child's SES quintile:

$$Y_{ij1} = -1.2 + 0.22 * [sex_{ij} = 1] + 0.08 * age_{ij1} + \\ 0.01 * [SES_{ij1} = \text{quintile 2}] + 0.37 * [SES_{ij1} = \text{quintile 3}] + \\ 0.33 * [SES_{ij1} = \text{quintile 4}] + 0.65 * [SES_{ij1} = \text{quintile 5}] + \psi_{ij}$$ (2)

where $\psi_{ij} \overset{iid}{\sim} N(0,1)$

v. Child's depression status ($depression_{ijk}$) at waves $k = 2, 4$ and 6 was generated using a GLMM conditional on child's age at wave 1, child's sex, NAPLAN scores at wave 1, child's SES quintile and wave:

$$logit\{P(depression_{ijk} = 1)\} = -4.0 + 0.31 * age_{ij1} + 0.08 * k + \\ (-0.52) * [sex_{ij} = 1] + (-0.05) * Y_{ij1} + \\ (-0.3) * [SES_{ij1} = \text{quintile 2}] + (-0.4) * [SES_{ij1} = \text{quintile 3}] + \\ (-0.57) * [SES_{ij1} = \text{quintile 4}] + (-0.86) * [SES_{ij1} = \text{quintile 5}] + u_{0i} + u_{0ij}$$ (3)

where $u_{0ij}$ and $u_{0i}$ are distributed as $u_{0ij} \overset{iid}{\sim} N(0, 1.5^2)$, $u_{0i} \overset{iid}{\sim} N(0, 0.25^2)$ respectively.

vi. We then generated the auxiliary variable, child's behavioural problems at waves 2, 4 and 6 ($SDQ_{ijk}$), which is not included in the analysis model but is associated with the exposure, using a LMM conditional on depression symptoms at waves 2, 4 and 6 and wave:

$$SDQ_{ijk} = 16 + 1.6 * depression_{ijk} + +(-0.1) * wave_{ijk} + \nu_{0i} + \nu_{0ij} + \epsilon_{ijk}$$ (6)

where $\epsilon_{ijk}$, $\nu_{0i}$ and $\nu_{0ij}$ are iid as; $\epsilon_{ijk} \sim N(0, 3.0^2)$ $\nu_{0i} \sim N(0, 0.8^2)$, and $\nu_{0ij} \sim N(0, 4.0^2)$ respectively.

vii. Finally child's standardized NAPLAN score($Y_{ijk}$) at waves $k = 3, 5$ and 7 was generated from a LMM as shown below:

$$Y_{ijk} = 2.0 + (-0.02) * depression_{ij(k-1)} + \\ (-0.01) * k + (-0.2) * age_{ij1} + 0.15 * [sex_{ij} = 1] + \\ 0.7 * Y_{ij1} + (-0.02) * [SES_{ij1} = \text{quintile 2}] + \\ (-0.10) * [SES_{ij1} = \text{quintile 3}] + (0.02) * [SES_{ij1} = \text{quintile 4}] +$$ (4)

$$(-0.02) * [SES_{ij1} = \text{quintile } 5] + (-0.01) * SDQ_{ijk} + a_{0i} + a_{0ij} + \varepsilon_{ijk}$$

where $\varepsilon_{ijk} \overset{iid}{\sim} N(0.25^2)$ and $a_{0i}$ and $a_{0ij}$ are distributed as $a_{0i} \overset{iid}{\sim} N(0, 0.05^2)$, $a_{0ij} \overset{iid}{\sim} N(0, 0.25^2)$ respectively.

## Simulation of missing data

To simulate missingness, data were set to missing in depressive symptom scores at waves 2, 4 and 6 (the exposure of interest), NAPLAN scores at wave 3,5 and 7 (the outcome), SDQ values at waves 2,4 and 6, SES and NAPLAN scores at baseline (wave 1), as detailed below.

i. Missing values in SES and NAPLAN scores at baseline were generated by drawing from a logistic regression model dependent on age at baseline and sex as shown below:

$$\text{logit}\{P(M_{SES_{ij1}} = 1)\} = -1.5 + 0.03 * age_{ij1} + 0.01 * sex_{ij} \tag{5}$$

$$\text{logit}\{P(M_{Y_{ij1}} = 1)\} = -2.1 + 0.05 * age_{ij1} + 0.02 * sex_{ij} \tag{6}$$

ii. Missing values in depressive symptom scores at waves 2, 4 and 6 were generated by drawing from a GLMM dependent on baseline variables age, wavesex, NAPLAN at wave 1, SES at wave 1, SDQ values (at waves 2,4 and 6), and NAPLAN scores (at waves 3,5 and 7)

$$\text{Logit}\{P(M_{depression_{ijk}} = 1)\} = -8.0 + 0.72 * age_{ij1} + (-0.11) * k + \tag{7}$$
$$(0.16) * [sex_{ij} = 1] + (-0.17) * Y_{ij1} +$$
$$(-0.39) * [SES_{ij1} = \text{quintile } 2] + (0.27) * [SES_{ij1} = \text{quintile } 3] +$$
$$(0.19) * [SES_{ij1} = \text{quintile } 4] + (-0.03) * [SES_{ij1} = \text{quintile } 5] +$$
$$(-0.13) * Y_{ij(k+1)} + +0.04 * SDQ_{ijk} + u_{0i} + u_{0ij}$$

where $u_{0ij}$ and $u_{0i}$ are distributed as $u_{0ij} \overset{iid}{\sim} N(0, 0.05^2)$, $u_{0i} \overset{iid}{\sim} N(0, 0.01^2)$ respectively.

iii. Missing values in NAPLAN scores at waves 3,5 and 7 were generated by drawing from a GLMM dependent on baseline variables age, wave, sex, NAPLAN at wave 1, SES at wave 1, SDQ values (at waves 2,4 and 6), and depression scores (at waves 2,4,6)

$$\text{Logit}\{P(M_{NAPLAN_{ijk}} = 1)\} = -23 + 1.77 * age_{ij1} + 0.7 * k + 0.01 * \tag{8}$$
$$[sex_{ij} = 1] + (-0.70) * Y_{ij1} + (-4.9) * [SES_{ij1} = \text{quintile } 2] + (-1.9) *$$
$$[SES_{ij1} = \text{quintile } 3] +$$
$$(2.19) * [SES_{ij1} = \text{quintile } 4] + (-2.35) * [SES_{ij1} = \text{quintile } 5] +$$
$$(-0.25) * depression_{ij(k-1)} + +0.11 * SDQ_{ij(k-1)} + u_{0i} + u_{0ij}$$

where $u_{0ij}$ and $u_{0i}$ are distributed as $u_{0ij} \overset{iid}{\sim} N(0, 2.0^2)$, $u_{0i} \overset{iid}{\sim} N(0, 0.4^2)$ respectively.

*S4 Sub-functions within jomo() wrapper function in the R package jomo*

| Higher-level wrappers | | jomo | | | |
|---|---|---|---|---|---|
| | Type of variables | jomo1 | jomo1ran | | jomo2 |
| Description | | MVNI<br><br>Not tailored for clustered data | MVNI<br>+ Random effects<br>+ No missing level 2 variables<br>+ Fixed cluster-specific level-1 covariance | MVNI<br>+ Random effects<br>+ No missing level 2 variables<br>+ Fixed/Random cluster-specific level-1 covariance | MVNI<br>+ Random effects<br>+ Missing level 2 variables<br>+ Common/Fixed/Random cluster-specific level-1 covariance |
| Sub-functions | Binary/categorical | jomo1cat | jomo1rancat | jomo1rancathr | jomo2com |
| | Continuous | jomo1con | jomo1rancon | jomo1ranconhr | jomo2hr |
| | Mixed | jomo1mix | jomo1ranmix | jomo1ranmixhr | |

Table 1: Summary of sub functions used by the jomo() wrapper function, given the type of model specification (e.g., not clustered or clustered) and the variable type being imputed

*S5: Results and computational time for MI methods*

MI approaches for longitudinal data

Point estimate (and standard error) for the effect of early depressive symptoms on subsequent standardized NAPLAN numeracy scores and the estimates for the variance components (VC).

***Using R***

| Method | Regression coefficient (SE) | VC- individual level | Residual error | Computational time (minutes) |
|---|---|---|---|---|
| JM-1L-wide | -0.005 (0.034) | 0.284 | 0.261 | 36.96 |
| FCS-1L-wide | -0.008 (0.033) | 0.280 | 0.262 | 1.52 |
| FCS-1Lwide-MTW | -0.012 (0.034) | 0.327 | 0.285 | 1.37 |
| JM-2L | -0.039 (0.052) | 1.199 | 0.585 | 14.35 |
| FCS-2L | -0.010 (0.034) | 0.295 | 0.264 | 14.92 |

***Using Stata***

| Method | Regression coefficient (SE) | VC- individual level | Residual error | Computational time (minutes) |
|---|---|---|---|---|

| Method | Regression coefficient (SE) | VC-school level | VC-individual level | Residual error | Computational time (minutes) |
|---|---|---|---|---|---|
| JM-1L-wide | -0.002 (0.032) | | 0.283 | 0.260 | 1.62 |
| FCS-1L-wide | -0.007 (0.034) | | 0.280 | 0.261 | 3.36 |
| FCS-1Lwide-MTW | -0.013 (0.036) | | 0.273 | 0.275 | 3.79 |

MI approaches for longitudinal data with higher-level clustering

Point estimate (and standard error) for the effect of early depressive symptoms on subsequent standardized NAPLAN numeracy scores and the estimates for the variance components (VC).

*Using R*

| Method | Regression coefficient (SE) | VC-school level | VC-individual level | Residual error | Computational time (minutes) |
|---|---|---|---|---|---|
| JM-1L-DI-wide* | -0.017 (0.034) | 0.056 | 0.236 | 0.267 | 63.39 |
| FCS-1L-DI-wide* | -0.012(0.034) | 0.055 | 0.239 | 0.266 | 4.37 |
| JM-2L-wide | -0.008 (0.037) | 0.048 | 0.239 | 0.272 | 55.91 |
| FCS-2L-wide* | -0.007 (0.038) | 0.048 | 0.239 | 0.264 | 636.0 |
| JM-2L-DI* | -0.146 (0.041) | 0.057 | 0.435 | 0.359 | 290.67 |
| FCS-2L-DI** | - | - | - | - | - |
| FCS-3L+ | -0.015(0.034) | 0.046 | 0.241 | 0.260 | 3.29 |

* The imputation model converged, post imputation analysis in one of the imputed datasets generated non-convergence warnings.
** The imputation model didn't converge (terminated with an error)
+ Singular fit warnings were generated for 13 (out of 660) imputation iterations, post imputation analysis in one of the imputed datasets did not achieve convergence.

*Using Stata*

| Method | Regression coefficient (SE) | VC-school level | VC-individual level | Residual error | Computational time (minutes) |
|---|---|---|---|---|---|
| JM-1L-DI-wide | -0.007(0.031) | 0.053 | 0.236 | 0.266 | 3.65 |
| FCS-1L-DI-wide | -0.013 (0.035) | 0.055 | 0.236 | 0.267 | 5.55 |